\documentclass[prl,preprint,showpacs,superscriptaddress,nofootinbib,floatfix]{revtex4}
\usepackage{latexsym}
\usepackage{epsfig}

\newcommand{\beq}{\begin{equation}}
\newcommand{\eeq}{\end{equation}}
\newcommand{\bea}{\begin{eqnarray}}
\newcommand{\eea}{\end{eqnarray}}

\begin{document}

\preprint{ANL-HEP-PR-03-073}
\preprint{FERMILAB-Pub-03/288-T}

\title{Lower limits on $R$-parity-violating couplings in supersymmetric
models with light squarks}

\author{Edmond~L.~Berger}
\affiliation{High Energy Physics Division, Argonne National Laboratory,
Argonne, IL 60439}
\author{Zack~Sullivan}
\affiliation{Theoretical Physics Department, Fermi National
Accelerator Laboratory,\\ Batavia, IL, 60510-0500}

\date{September 30, 2003}

\begin{abstract}
We interpret the results of searches for strongly interacting massive
particles to place absolute lower limits on $R$-parity-violating
couplings for squarks with mass ($m_{\tilde{q}}$) below 100 GeV. Recent
searches for anomalous isotopes require that there be a
baryon-number-violating or lepton-number-violating coupling larger
than $10^{-22}$--$10^{-21}$ if $m_{\tilde{q}}>18$ GeV.  Using data
from searches for stable particles at the CERN Large Electron Positron
collider (LEP) we demonstrate that this lower limit increases by 14
orders of magnitude, to an $R$-parity-violating coupling larger than
$10^{-8}$--$10^{-7}$ for any squarks of mass less than 90 GeV.  In the
presence of an $R$-parity-violating coupling of this magnitude,
neutralinos cannot explain the dark matter density in the Universe.
\end{abstract}


\maketitle

\label{sec:intro}

In supersymmetric extensions of the standard model of elementary
particle physics, particles are typically assigned a new quantum
number called $R$ parity ($R_p$)~\cite{Fayet:1974pd}.  Particles of
the standard model are defined to have even $R_p$, and their
corresponding superpartners have odd $R_p$. If $R$ parity is
conserved, then superpartners must be produced in pairs, each of which
decays to a final state that includes at least one stable lightest
supersymmetric particle (LSP).  One appealing feature of $R$-parity
conservation is that if the LSP is a neutralino, then it is naturally
predicted to exist with the necessary relic density to explain
the dark matter density in the Universe~\cite{Jungman:1995df}.

In contrast, if $R$ parity is not conserved, then supersymmetric
particles may decay into standard model particles, and supersymmetry
may not provide a explanation of dark matter.  Further,
many of the postulated signatures of supersymmetry (SUSY), such as
missing energy in high energy collider experiments, may not exist.  In
particular, many of the exclusion limits on squark masses may no
longer apply if $R$ parity is violated.  On the other hand, if the
$R$-parity-violating couplings are small enough, then superparticles
may appear to be stable on a time-scale relevant for collider
searches.  In the search for physics beyond the standard model in
cosmological and terrestrial experiments, it is of substantial
importance to address explicitly the question of $R$-parity violation,
a possibility that we explore in this Letter.

The present \emph{upper} bounds on possible $R$-parity-violating couplings
are obtained principally from next-to-leading order quantum
corrections to the rates for particle decays and neutral meson mixing.
These bounds are relatively restrictive for the first generation of
quarks and leptons but are much less so for states of the second and
third generations \cite{Allanach:1999bf}.  In this Letter, we approach
$R$-parity-violating couplings from the other direction.  We
demonstrate that there must be strong \emph{lower} bounds on
$R$-parity-violating couplings for a large range of squark masses.
Otherwise, squarks with mass in these ranges cannot exist.

We focus on squarks with mass below 100~GeV. Although it is widely
held that squarks in this range are ruled out by experiment, this
limitation does not apply in general SUSY-breaking scenarios if the
states with lowest mass are coupled feebly to the neutral gauge boson
$Z$.  One such scenario \cite{Berger:2000mp} involving a light bottom
squark $\widetilde{b}$ and a gluino of moderate mass provides an
explanation for the large excess in the $b$-quark cross section
measured at hadron colliders \cite{bexcess,caveat}, and is supported
by the recent observation of a large time-averaged $B^0 \bar{B}^0$
mixing probability $\bar{\chi}$~\cite{Acosta:2003ie}.  If the lighter
of the two bottom squarks is an appropriate mixture of left-handed and
right-handed squarks, its tree-level coupling to the $Z$ can be
arbitrarily small, permitting good agreement with the precise
measurements of $Z$-peak observables~\cite{Carena:2000ka}, including
effects at loop-level~\cite{Berger:2003sn}.

In this Letter we go beyond Ref.\ \cite{Berger:2000mp}, and examine
the decays of all flavors of squarks.  We demonstrate that a lower
bound on baryon-number or lepton-number violating couplings of
$10^{-22}$--$10^{-21}$ can be extracted from primordial abundance
searches.  Then, using data from searches for massive stable particles
produced at LEP, we place absolute lower bounds on
$R$-parity-violating couplings as a function of squark mass.  For any
squarks with mass less than 90~GeV, we show that the lower bounds on
baryon-number or lepton-number violating couplings improve to
$10^{-8}$--$10^{-7}$.

\section{Supersymmetric notation}
\label{sec:note}

There are several popular choices of convention for the squark 
mixing angles and for the normalization of
$R$-parity-violating couplings.  For evaluating the
lifetimes of squarks, we use the $R$-parity-violating 
superpotential~\cite{Weinberg:1981wj}
\beq
W_{\slash\!\!\!\!{R_p}}=\epsilon^{\sigma\rho}\lambda^\prime_{ijk}
L_{i\sigma}Q^\alpha_{j\rho}D^c_{k\alpha} + 
\epsilon^{\alpha\beta\gamma}\lambda^{\prime\prime}_{ijk}
U^c_{i\alpha}D^c_{j\beta}D^c_{k\gamma} \,,
\eeq
where ($\sigma$, $\rho$) are $SU(2)$ indices, ($\alpha$, $\beta$,
$\gamma$) are $SU(3)$ color indices, and ($i$, $j$, $k$) are
generation indices. $L$ and $Q$ are leptonic and hadronic left-handed
superdoublets, respectively.  $U^c$ and $D^c$ are right-handed up- and
down-type conjugate superfields, respectively.  The first term
violates lepton number, and the second term violates baryon number.
The dimensionless coupling strengths $\lambda^\prime_{ijk}$ and
$\lambda^{\prime\prime}_{ijk}$ measure the strength of lepton-number
and baryon-number violation, respectively.  In general the couplings
are complex.

The physical squarks are a mixture of the scalar partners of the left-
and right-chiral quarks.  The mass eigenstates are two complex
scalars, $\tilde q_1$ and $\tilde q_2$, expressed in terms of
left-handed (L) and right-handed (R) squarks, $\tilde{q}_L$ and
$\tilde{q}_R$.  We choose the squark mixing matrix such that
$\tilde{q}_1$ is mostly right-handed when the mixing angle is
small.\footnote{Note we use a different mixing convention here than
presented in Ref.~\protect\cite{Berger:2000zk}.}
\beq
\left( \begin{array}{c} \tilde{q}_1 \\ \tilde{q}_2 \end{array} \right)
= \left( \begin{array}{rr} \cos\theta_{\tilde{q}} & \sin\theta_{\tilde{q}} \\
-\sin\theta_{\tilde{q}} & \cos\theta_{\tilde{q}} \end{array} \right)
\left( \begin{array}{c} \tilde{q}_R \\ \tilde{q}_L \end{array} \right) \,.
\eeq

Given the mixing matrix above, the partial widths of a squark decaying
into two massless quarks are
\bea
\Gamma(\tilde{q}_1\to q q) &=& \frac{m_{\tilde{q}_1}}{2\pi} \cos^2
\theta_{\tilde{q}} \, |\lambda^{\prime\prime}_{ijk}|^2 \,,\nonumber \\
\Gamma(\tilde{q}_2\to q q) &=& \frac{m_{\tilde{q}_2}}{2\pi} \sin^2
\theta_{\tilde{q}} \, |\lambda^{\prime\prime}_{ijk}|^2 \,, \, j<k \,.
\label{eq.:bviol}
\eea
The partial widths of an up-type squark that decays into a massless quark and
lepton are 
\bea
\Gamma(\tilde{u}_{1j}\to \overline{e}_i d_k) &=& \frac{m_{\tilde{u}_{1j}}}{16
\pi} \sin^2 \theta_{\tilde{u}_j} |\lambda^{\prime}_{ijk}|^2 \,, \nonumber \\
\Gamma(\tilde{u}_{2j}\to \overline{e}_i d_k) &=& \frac{m_{\tilde{u}_{2j}}}{16
\pi} \cos^2 \theta_{\tilde{u}_j} |\lambda^{\prime}_{ijk}|^2 \,,
\label{eq.:lviolup}
\eea
and for a down-type squark, they are 
\bea
\Gamma(\tilde{d}_{1j}\to \overline{\nu}_i d_k) &=& \frac{m_{\tilde{d}_{1j}}}{16
\pi} \sin^2 \theta_{\tilde{d}_j} |\lambda^{\prime}_{ijk}|^2 \,, \nonumber \\
\Gamma(\tilde{d}_{2j}\to \overline{\nu}_i d_k) &=& \frac{m_{\tilde{d}_{2j}}}{16
\pi} \cos^2 \theta_{\tilde{d}_j} |\lambda^{\prime}_{ijk}|^2 \,, \\
\Gamma(\tilde{d}_{1k}\to \nu_i d_j ) &=& \frac{m_{\tilde{d}_{1k}}}{16\pi}
\cos^2 \theta_{\tilde{d}_k} |\lambda^{\prime}_{ijk}|^2 \,, \nonumber \\
\Gamma(\tilde{d}_{2k}\to \nu_i d_j ) &=& \frac{m_{\tilde{d}_{2k}}}{16\pi}
\sin^2 \theta_{\tilde{d}_k} |\lambda^{\prime}_{ijk}|^2 \,, \\
\Gamma(\tilde{d}_{1k}\to e_i u_j ) &=& \frac{m_{\tilde{d}_{1k}}}{16\pi} \cos^2
\theta_{\tilde{d}_k} |\lambda^{\prime}_{ijk}|^2 \,, \nonumber \\
\Gamma(\tilde{d}_{2k}\to e_i u_j ) &=& \frac{m_{\tilde{d}_{2k}}}{16\pi} \sin^2
\theta_{\tilde{d}_k} |\lambda^{\prime}_{ijk}|^2 \,.
\label{eq.:lvioldn}
\eea

If one $R$-parity-violating coupling is large compared to the others,
then only one of these partial widths will be applicable at any given
time.  For simplicity we present results assuming one significant
coupling.  Any other combination of couplings may be derived using the
widths above.

\section{Limits on squark masses and $R$-parity-conserving couplings}
\label{subsec:rpc}

To place $R$-parity-\emph{violating} couplings into context, we begin
with a brief summary of the current limits on
$R$-parity-\emph{conserving} decays of squarks.  In
$R$-parity-conserving SUSY, squarks are either completely stable, or
undergo $R$-parity-conserving decays into final states containing a
stable sparticle.  If the daughter sparticle is neutral, then there
will be missing energy in the event.  Many limits on squark masses are
based on searches for jets plus missing energy, and some are
predicated on the assumption that couplings to the $Z$ boson are not
suppressed.  In our work, by contrast, we wish to place the most
conservative lower limits on $R$-parity-violating couplings, and we 
therefore include cases in which the squarks decouple from the
$Z$ boson.

In the absence of strong coupling to the $Z$ boson, direct limits on
the existence of up-type and down-type squarks differ significantly.
Because of their larger coupling to photons, the $R$-parity-conserving
decay of up-type squarks to any lightest supersymmetric particle (LSP)
is ruled out for $m_{\tilde{q}}<63$ GeV
\cite{Heister:2003hc,Hagiwara:fs,Naroska:1986si,Barate:2000tu}
as long as the lifetime of the squark is less than 1--10 ns
\cite{Heister:2003hc}.
This limit increases to 83--89 GeV in any case where $\Delta m =
m_{\tilde{q}} - m_{\mathrm{LSP}} > 5$ GeV
\cite{Barate:2000tu,Abbiendi:1999yz}.  Down-type squarks with masses
of 40--89 GeV, $R$-parity-conserving decays, and $\Delta m > 5$
GeV are also excluded.  However, below 40 GeV the only limits on
$R$-parity-conserving decays for down-type squarks come from
exclusions of possible decay products.  All sleptons, charginos and
neutralinos are excluded below 37 GeV \cite{Hagiwara:fs}, and stable
gluinos are excluded below 27 GeV \cite{Heister:2003hc}.  Unless there
is a stable gluino of mass 27--35 GeV, or finely-tuned LSP mass, all
squarks less than 83--89 GeV will either be appear to be stable on the
time-scale of accelerator experiments, or decay via
$R$-parity-violating interactions.  For the rest of this Letter, we will
consider the $R$-parity-conserving decays to be excluded.

\section{$R$-parity-violating couplings}

Limits on the relic abundances of stable primordial particles may be
used to set an initial lower bound on $R$-parity-violating couplings.
For hadron masses less than 100 GeV, Fermi motion in nuclei is
expected to prevent the capture of exotic states in the nuclei of
light elements, such as sodium or oxygen.  Heavy elements, such as
gold and iron, are believed to be capable of capturing the strongly
interacting massive particles (SIMPs)~\cite{Mohapatra:1999gg}.
Recently, a search for anomalous gold and iron isotopes was used to
set limits on the cosmological density parameter $\Omega_S$, the ratio
of the density of relic primordial SIMPs to the critical density.
Upper limits of $\Omega_S \approx 6\times 10^{-8}(m_{\tilde{q}}
[\textrm{GeV}])^{1.7}$ were extracted for the mass range 2.8 to
100~GeV~\cite{Javorsek:ASTROJ}.
Following Ref.~\cite{Kolb}, we calculate that the relic abundance of
squarks, including all QCD and supersymmetric annihilations,
contributes approximately $10^{-8} (m_{\tilde{q}}
[\textrm{GeV}])^{2.29}$ to $\Omega_S$.  We find a freeze-out
temperature that varies between $T_F=m_{\tilde{q}}/36$ and
$m_{\tilde{q}}/30$.  The QCD annihilation into 2 gluons completely
dominates the annihilation rate.  Hence, the abundance of stable
squarks depends solely on the mass of the squarks, and no other
supersymmetric parameters.  If the assumptions regarding exposure
times for the samples and capture cross sections are correct, then
completely stable primordial squarks may be excluded between 18--100
GeV.

To evade the primordial abundance bounds the squarks must have decayed
with a lifetime less than the age of the Universe.  For reasons
discussed above we assume that $R$-parity-conserving decays are
excluded for squarks of mass less than $100$~GeV.  Using our complete
relic density calculation, Eqs.\
(\ref{eq.:bviol})--(\ref{eq.:lvioldn}), and exact limits on $\Omega_S$
from Ref.\ \cite{Javorsek:ASTROJ}, we conclude that there must be at
least one $R$-parity-violating coupling $\lambda^{\prime\prime} >
(2$--$3)\times 10^{-22}$ or $\lambda^{\prime} > (6$--$9)\times
10^{-22}$ if $20<m_{\tilde{q}}<100$ GeV.  In the case of mixing, this
limit is larger by $|1/\sin \theta_{\tilde{q}}|$ or
$|1/\cos\theta_{\tilde{q}}|$.

While a few combinations of masses and $R$-parity-violating couplings
have been excluded \cite{Savinov:2000jm}, we now show that there is a
strong lower limit on at least one baryon-number or lepton-number
violating coupling for all squarks of mass less than 85--90 GeV by
using the results of a new ALEPH study~\cite{Heister:2003hc,Janotcom}.
From the absence of heavy hadron tracks and unexplained missing
energy, the ALEPH study concludes that squarks, if produced, decay
before leaving a discernible signature.  That investigation rules out
squarks that live more than a few nanoseconds between the electroweak
precision lower limit of 1.3~GeV and 92 (95) GeV for down-type
(up-type) squarks~\cite{Heister:2003hc,Janotcom}.  Working from the
results of that paper, we use our
Eqs.\ (\ref{eq.:bviol})--(\ref{eq.:lvioldn}) to derive the minimal
necessary $R$-parity-violating couplings under the assumption of only
one non-zero coupling.

\begin{figure}[tbh]
\begin{center}
\epsfxsize=4.5in \epsfbox{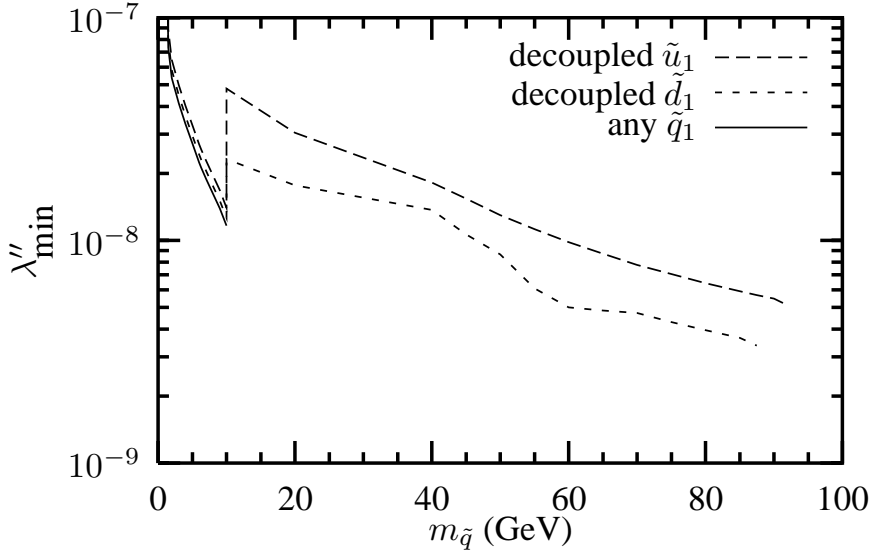}
\end{center}
\caption{Lower limits on the baryon-number-violating couplings
$\lambda^{\prime\prime}$ as a function of squark $\tilde{q}_1$ mass.
The solid line ($m_{\tilde{q}}<10$ GeV) indicates an absolute lower
limit on the couplings.  The dashed lines indicate the limits where
$\tilde{d}_1$ or $\tilde{u}_1$ decouple from the $Z$-boson ($\sin
\theta_{\tilde{d}} \approx 0.39$ or $\sin \theta_{\tilde{u}} \approx 0.56$).
For $m_{\tilde{q}}>10$ GeV the dashed lines are also absolute lower limits.
\label{fig:lims}}
\end{figure}

\begin{figure}[tbh]
\begin{center}
\epsfxsize=4.5in \epsfbox{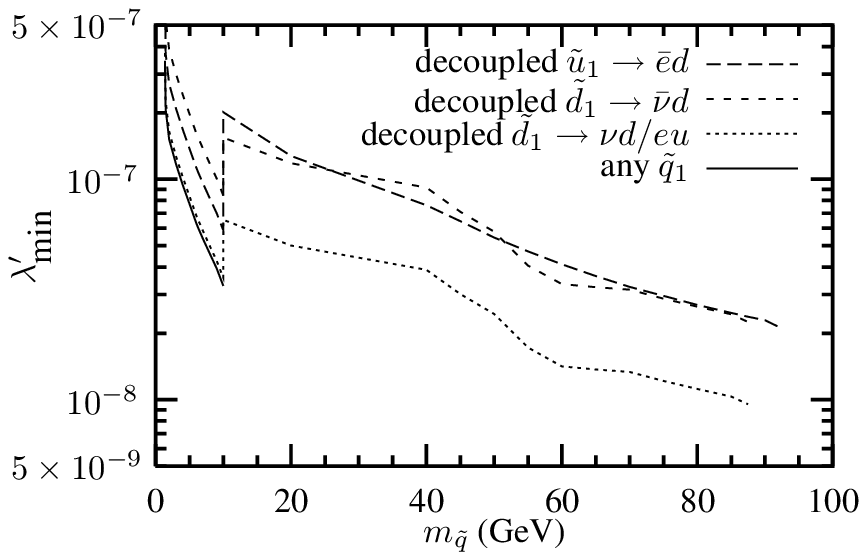}
\end{center}
\caption{Lower limits on the lepton-number violating couplings
$\lambda^{\prime}$ as a function of squark mass.  The solid line
($m_{\tilde{q}}<10$ GeV) indicates an absolute lower limit on the
couplings.  The dashed lines indicate the limits where $\tilde{d}_1$
or $\tilde{u}_1$ decouple from the $Z$-boson ($\sin \theta_{\tilde{d}}
\approx 0.39$ or $\sin \theta_{\tilde{u}} \approx 0.56$), and are listed
by decay.  For $m_{\tilde{q}}>10$ GeV the dashed lines are also
absolute lower limits.
\label{fig:lims2}}
\end{figure}

We present our results in Figs.~\ref{fig:lims} and~\ref{fig:lims2}.
The curves in Fig.~\ref{fig:lims} show that there must be a
baryon-number-violating coupling $\lambda^{\prime\prime} \agt
5 \times 10^{-9}$--$10^{-7}$ for squark masses less than 90
GeV.  If there are no large baryon-number-violating couplings, then
the curves in Fig.~\ref{fig:lims2} show that there must be a
lepton-number-violating coupling $\lambda^{\prime} \agt
10^{-8}$--$5 \times 10^{-7}$ for squark masses less than 90 GeV.  In
these figures, we distinguish the minimal couplings obtained in the
cases where the lightest squarks ($\tilde{d}_1$ or $\tilde{u}_1$)
decouple from the $Z$-boson ($\sin \theta_{\tilde{d}} \approx 0.39$ or
$\sin \theta_{\tilde{u}} \approx 0.56$).  The decrease with mass of
the lower limits in part reflects the fact that partial widths in
Eqs.\ (\ref{eq.:bviol})--(\ref{eq.:lvioldn}) are proportional to the squark
mass.

There are two distinct regions in Figs.~\ref{fig:lims}
and~\ref{fig:lims2}.  In the region of small squark masses, the
assumed production process is $e^+e^- \rightarrow q \bar{q} g$, with
$g \rightarrow \tilde{q} \tilde{q}^\ast$.  Beyond a squark mass of
roughly 10 GeV, there are not enough events to provide a competitive
limit.  In the higher mass region the assumed production process is
$\tilde{q} \tilde{q}^\ast$ pair production, and a lifetime limit of
about 1 ns can be set up to 92 GeV.  Below a squark mass of 10 GeV the
relevant production process is independent of mixing angle.
Therefore, the lower bound for a given squark mixing-angle is $|1/\sin
\theta_{\tilde{q}}|$ or $|1/\cos \theta_{\tilde{q}}|$ larger than the
absolute bound.  Above 10 GeV the cross section for pair-production is
measured under the assumption that the squarks are decoupled from the
$Z$ boson.  If the mixing angle changes, then the squarks couple to
the $Z$-boson and the cross section used to estimate the lifetime
increases dramatically.  The mixing-angle dependence will temper or
enhance the large increase in the lower bound in a
experiment-dependent manner that we do not attempt to estimate.

\section{Conclusions}
\label{sec:concl}

If there are squarks with mass less than about 90 GeV, both exotic
isotope searches and data from LEP require that they are not stable.
Since $R$-parity-conserving decay modes are explicitly excluded,
exotic isotope searches imply that $R$-parity-violating couplings
greater than $10^{-22}$--$10^{-21}$ must be present if there are
squarks of mass 18--100 GeV.  The data from LEP improve this
lower bound by 14 orders of magnitude, and require there to be a
baryon-number or lepton-number violating coupling greater than
$10^{-8}$--$10^{-7}$ for squarks of mass less than 90 GeV.  These
lower limits can be significantly larger if there is large mixing in
the squark sector or non-vanishing coupling to the $Z$ boson.

In the class of supersymmetric models we consider, the LSP can be a
neutralino as long as the partial width for the decay of the lightest
squark into a neutralino is smaller than the partial width for the
$R$-parity-violating decay of the squark.  However, the existence of
$R$-parity-violating couplings means that a neutralino LSP will decay
through an off-shell squark to standard-model particles.  A minimal
$R$-parity-violating coupling of $10^{-22}$, as allowed by exotic
isotope searches, is not large enough to significantly reduce the
number density of relic neutralinos.  However, the data from LEP imply
a lifetime of neutralinos that is $10^{28}$ times shorter, meaning
that the neutralinos will have a negligible relic density.  Hence,
neutralinos cannot explain the excess of dark matter in the Universe
in the presence of squarks with mass less than 90 GeV.

When combined with upper limits on $R$-parity-violating couplings, our
study leads to the conclusion that there may be very interesting
signals of $R$-parity-violating supersymmetry at high energy colliders.  
In particular, squarks may be produced with lifetimes that are long
enough to produce displaced vertices when they decay.  In addition,
$R$-parity-violation offers the possibility that superpartners may be
produced singly, rather than in pairs, a subprocess that has
significant advantages especially at collider energies that are not
substantially higher than the masses of the produced
states~\cite{Berger:2000zk}. We urge additional attention in current
and anticipated experiments to searches for explicit
$R$-parity-violating decay modes.


We acknowledge valuable communications with P.~Janot, CERN.  This work
was supported by the U.~S.~Department of Energy under Contract Nos.\
W-31-109-ENG-38 and DE-AC02-76CH03000.


\end{document}